\documentclass[a4paper,UKenglish,autoref]{lipics-v2019}

\usepackage[ruled,vlined,linesnumbered,commentsnumbered]{algorithm2e}
\usepackage{amsmath,amssymb,amsfonts}

\title{Data structures for connectivity and cut queries}
\titlerunning{Data structures for connectivity and cut queries}

\author{Zeev Nutov}{The Open University of Israel}{nutov@openu.ac.il}
{https://orcid.org/0000-0002-6629-3243}{}
\authorrunning{Zeev Nutov}
\Copyright{Zeev Nutov}

\nolinenumbers


\acknowledgements{}

\begin{document}

\maketitle

\newcommand{\sem}    {\setminus}
\newcommand{\subs}   {\subseteq}
\newcommand{\empt}  {\emptyset}

\newcommand{\f}   {\frac}

\newcommand{\Aa}  {{\cal A}}
\newcommand{\Bb}  {{\cal B}}
\newcommand{\CC}  {{\cal C}}
\newcommand{\FF}  {{\cal F}}
\newcommand{\HH}  {{\cal H}}
\newcommand{\PP}  {{\cal P}}
\newcommand{\RR}  {{\cal R}}
\newcommand{\Sa}   {{\cal S}}
\newcommand{\TT}   {{\cal T}}
\newcommand{\LL}   {{\cal L}}

\newcommand{\con}   {{\sc pcon$_k(s,t)$}} 
\newcommand{\cut}     {{\sc pcut$_k(s,t)$}} 
\newcommand{\econ} {{\sc con$_k(s,t)$}}
\newcommand{\mcut}  {{\sc cut$_k(s,t)$}}  

\newcommand{\ga}   {\gamma}
\newcommand{\de}   {\delta}
\newcommand{\ka}   {\kappa}
\newcommand{\la}   {\lambda}

\newcommand{\A}   {\mathbb{A}}
\newcommand{\B}   {\mathbb{B}}
\newcommand{\C}   {\mathbb{C}}
\newcommand{\p}   {\partial}

\def\lca   {{\sf lca}} 
\def\dist  {{\sf dist}} 
\def\o     {{\sf o}}
\newcommand{\Ga}  {\Gamma}

\newcommand {\ignore} [1] {}


\begin{abstract}
Let $\ka(s,t)$ denote the maximum number of internally disjoint $st$-paths in an undirected graph $G$.
We consider designing a compact data structure that answers $k$-bounded node connectivity queries:
given $s,t \in V$ return $\min\{\ka(s,t),k+1\}$. 
A trivial data structure has space $O(n^2)$ and query time~$O(1)$.
A data structure of Hsu \& Lu \cite{HL} has space $O(k^2n)$ and query time $O(\log k)$,
and a randomized data structure of Iszak \& Nutov \cite{IN} has space $O(kn\log n)$ and query time $O(k \log n)$.
We extend the Hsu-Lu data structure to answer queries in time $O(1)$. 
In parallel to our work, Pettie, Saranurak \& Yin \cite{PSY} extended the Iszak-Nutov \cite{IN} data structure 
to answer queries in time $O(\log n)$. 
Our data structure is more compact than that of \cite{PSY} for $k<\log n$, and our query time is always better.

We then augment our data structure by a list of cuts that enables to return 
a pointer to a minimum $st$-cut in the list (or to a cut of size $\leq k$) whenever $\ka(s,t) \leq k$.
A trivial data structure has cut list size $n(n-1)/2$, and cut query time $O(1)$, while the 
Pettie, Saranurak \& Yin \cite{PSY} data structure has list size $O(kn \log n)$ and cut query time $O(\log n)$.
We show that $O(kn)$ cuts suffice to return an $st$-cut of size $\leq k$,
and a list of $O(k^2 n)$ cuts contains a minimum $st$-cut for every $s,t \in V$.

In the case when $S$ is a node subset with $\ka(s,t) \geq k$ for all $s,t \in V$, 
we show that $3|S|$ cuts suffice, and that these cuts can be partitioned into $O(k)$ laminar families.
Thus using space $O(kn)$ we can answers each connectivity and cut queries for $s,t \in S$ in $O(1)$ time,
generalizing and substantially simplifying the proof of a result of Pettie and Yin \cite{PY} for the case $|S|=V$. 
\end{abstract}

\section{Introduction} \label{s:intro}

Let $\ka(s,t)=\ka_G(s,t)$ denote the maximum number of internally disjoint $st$-paths in a graph $G=(V,E)$.
An {\bf $st$-cut} is a subset $Q \subs V \cup E$ such that $G \sem Q$ has no $st$-path. 
By Menger's Theorem, $\ka(s,t)$ equals to the minimum size of an $st$-cut, and there always 
exists a minimum $st$-cut that contains no edge except of $st$.
We consider designing a compact data structure that given $s,t \in V$ and $k < n=|V|$ answers the following 
$k$-bounded connectivity/cut queries. 

\begin{tabbing}
{\con} \ \= (partial connectivity query): \ \ \= Determine whether $\ka(s,t) \leq k$.                                    \\
{\cut}    \> (partial cut query):                    \> If $\ka(s,t) \leq k$ then return an $st$-cut of size $\leq k$. \\
{\econ} \> (connectivity query):                \>             Return $\min\{\ka(s,t),k+1\}$.                                    \\
{\mcut} \> (min-cut query):                      \> If $\ka(s,t) \leq k$ then return a minimum $st$-cut.
\end{tabbing}

The cut queries {\cut} and {\mcut} require $\Theta(k)$ time just to write an $st$-cut.
It is therefore makes sense to allow the data structure to include a list of cuts,  
and to return just a pointer to an $st$-cut in the list. 
How short can this list be? By choosing a minimum $st$-cut for each pair $\{s,t\}$, 
one gets a list of $n(n-1)/2$ cuts. 
This gives a trivial data structure, that answers queries in $O(1)$ time,
but has $n(n-1)/2$ cuts -- just store the pairwise connectivities in an $n \times n$ matrix, 
with relevan pointers to cuts.
For edge connectivity, the Gomory-Hu Cut-Tree \cite{GH} 
shows that there exists such a list of $n-1$ cuts that form a laminar family.
However, no similar result is known for the node connectivity case considered here.

Hsu and Lu \cite{HL} showed that for any graph $G=(V,E)$ 
there exists an auxiliary graph $H=(V,F)$ 
and an ordered partition $\PP=(S_1,\ldots, S_q)$ of $V$, such that the following holds:
\begin{itemize}
\item
In $H$, every part $S_i$ has at most $2k-1$ neighbors in $S_{i+1} \cup \cdots \cup S_q$; hence $|F|=O(kn)$.
\item
$\ka(s,t) \geq k+1$ iff $s,t$ belong to the same part of $\PP$ or $st \in F$. 
\end{itemize}
They also gave a polynomial time algorithm for constructing such $H$ and $\PP$.
Augmenting $H$ by a perfect hashing data structure enables to answer ''$st \in F$?'' queries in $O(1)$ time.
Since $|F| =O(kn)$, this gives an $O(kn)$ 
space\footnote{As in previous works, we ignore the unavoidable $O(\log n)$ factor invoked by storing the indexes of nodes, 
and assume that any basic arithmetic or comparison operation with indexes can be done in $O(1)$ time.}
data structure that determines whether $\ka(s,t) \geq k+1$ in $O(1)$~time.
Furthermore, a collection of such data structures for each $k'=1, \ldots ,k+1$ has space $O(k^2 n)$ 
and enables to find $\min\{\ka(s,t),k+1\}$ in $O(\log k)$ time, using binary search. 
We improve the query time to $O(1)$. 

\begin{theorem} \label{t1}
There exists an $O(k^2 n)$ space data structure that answers {\econ} queries in $O(1)$ time.
\end{theorem}

Our data structure is easy to describe. 
Let $T=(V_T,E_T)$ be a tree rooted at $r$ with leaf set $V$
and integer {\bf levels} $\{\ell(v):v \in V_T\}$ such that $\ell(r)=0$ and $\ell(u)>\ell(v)$ if $u$ is a child of $v$;
we will call such a pair $\langle T,\ell \rangle$ a {\bf leveled tree}. 
Let $\lca(s,t)=\lca_T(s,t)$ denote the lowest common ancestor of $s,t$ in $T$. 
Our data structure for answering {\econ} queries is described in the following theorem. 

\begin{theorem}  \label{t2}
For any graph $G=(V,E)$ and a positive integer $k < n=|V|$ there exist a leveled 
tree $\langle T,\ell \rangle$ with leaf set $V$ and $O(n)$ edges, and
an edge weighted graph $\langle H=(V,F),w \rangle$ with $O(k^2 n)$ edges,
such that the following holds: 
\[
\min\{\ka(s,t),k+1\} = \left \{ \begin{array}{ll}
w(st)            & \ \mbox{if} \ \ st \in F \\
 \ell(\lca_T(s,t)) & \mbox{ otherwise}
\end{array} \right .
\]
\end{theorem}

Augmenting $T$ by Gabow \& Tarjan \cite{GT} lowest common ancestor data structure enables to find $\ell(\lca_T(u,v))$ in $O(1)$ time. 
Augmenting $H$ by a perfect hashing data structure enables in $O(1)$ time to return $w(st)$ or to determine that $st \notin F$. 
Since $|F| =O(k^2n)$, this gives an $O(k^2n)$ space data structure that answers {\econ} queries in $O(1)$~time,
as required in Theorem~\ref{t1}. 

In parallel to our work, Pettie, Saranurak, \& Yin \cite{PSY} 
extended the randomized $O(kn \log n)$ space data structure of Izsak \& Nutov \cite{IN}, 
that in turn is based on an idea of Chuzhoy and Khanna \cite{CK}. 
The Izsak \& Nutov data structure answer {\econ} queries in $O(k \log n)$ time,
and Pettie, Saranurak, \& Yin extended it to answer {\econ} queries in $O(\log n)$ time. 
We briefly describe these results.
Given a set $S \subs V$ of {\bf terminals}, the edges and the nodes in $V \sem S$ are called {\bf elements}.
The {\bf element connectivity} between $s,t \in S$ is the maximum number of 
pairwise element disjoint $st$-paths.
The Gomory-Hu tree extends to element connectivity (c.f. \cite{Sch,CRX}), and implies an $O(|S|)$
space data structure that answers element connectivity queries between terminals in $O(1)$ time. 
The  data structure of \cite{IN}  decomposes a node connectivity instance 
into $O(k^2 \log n)$ element connectivity instances with $\Theta(n/k)$ terminals each;
we will give a generalization of this decomposition in Section~\ref{s:elc}. 
For any element connectivity instance $G_S$ with terminal set $S$ and $s,t \in S$, 
$\ka(s,t)$ is at most the element $st$-connectivity in $G_S$, and for at least one instance an equality holds
(an instance with $s,t \in S$ and $Q \cap S =\empt$ for some minimum $st$-cut $Q$).
So, to find $\kappa(s,t)$, one has to find the minimum element $st$-connectivity, 
among all instances in which $s,t$ are terminals.
There are $O(k^2 \log n)$ element connectivity instances, 
with $\Theta(n/k)$ terminals each, hence the overall number of terminals is $O(k^2 \log n) \cdot (n/k)=O(kn \log n)$.
Iszak and Nutov \cite{IN} considered designing a 
labeling scheme\footnote{For additional work on labeling schemes for node connectivity, see for example, \cite{KKKP,Kor,HL}.}, 
where efficient query time is not required,
and used the connectivity classes data structure in each element connectivity instance;
this enables to answer {\econ} queries in $O(k \log n)$ time.
Pettie, Saranurak, and Yin \cite{PSY} used element connectivity Gomory-Hu trees instead,  
and also designed a novel data structure that given $s,t$ finds an instance with 
the minimum element $st$-connectivity in $O(\log n)$ time.
They also showed that for large values of $k$, any data structure for answering node connectivity queries 
needs at least $\Omega(kn/\log n)$ space, matching up to an $O(\log n)$ factor the $O(kn)$ space 
of a sparse certificate graph \cite{NI}.

Let us compare between the \cite{PSY} $O(kn \log n)$ space data structure to our results. 
Our query time is always better than that of \cite{PSY}  -- $O(1)$ vs. $O(\log n)$, and we use less space for $k<\log n$.  
Note that for $k \leq 4$ are known linear space data structures that can answer cut queries in $O(1)$ time;
see \cite{HT} and \cite{KTBC} for the cases $k=3$ and $k=4$, respectively. 
Our work, that was done independently from \cite{PSY} and uses totally different techniques,
extend this to any constant $k$, bridging the gap between the result of \cite{PSY} and 
the results known for $k \leq 4$. 
See Table~1 for comparison between some data structures.

\begin{table} [htbp] 
\begin{center}
\begin{tabular}{|c|c||c|c||c|}  \hline  
{\bf space}          & {\econ}          & {\bf list size}      & {\mcut}         & {\bf reference}     
\\\hline \hline
$O(n^2)$           & $O(1)$          & $n(n-1)/2$        & $O(1)$         & folklore
\\\hline 
$O(kn)$             & $T(n,k)$       & -                         & $T(n,k)$        & \cite{NI}      
\\\hline
 $O(k^2n)$        & $O(\log k)$   & -                         & -                    & \cite{HL} 
\\\hline 
$O(kn \log n)$  & $O(\log n)$  & $O(kn \log n)$  & $O(\log n)$  & \cite{PSY}
\\\hline
$O(k^2 n)$        & $O(1)$         & $O(k^2 n)$       & $O(1)$         & this paper
\\\hline
\end{tabular}
\end{center}
\caption{Summary of known data structures for connectivity and cut queries.
In the second row, $T(n,k)$ is the time needed to compute a minimum $st$-cut in a graph with $O(kn)$ edges. 
In the third row, the \cite{HL} data structure can answer {\con} query in $O(1)$ time using $O(kn)$ space.  
In the last row, our list size can be reduced to $O(kn)$ if we just want to answer {\cut} queries.}
\label{tbl:cut}
\end{table}

A graph is {\bf $k$-connected} if $\ka(s,t) \geq k$ for all $s,t \in V$.
Recently, Pettie and Yin \cite{PY}, and earlier in the 90's Cohen, Di Battista, Kanevsky, and Tamassia \cite{CDKT},
considered the above problem in $k$-connected graphs.
Pettie and Yin \cite{PY} suggested for $n \geq 4k$ an $O(kn)$ space data structure, 
that answers {\econ} in $O(1)$ time and {\mcut} in $O(k)$ time; 
they showed that it can be constructed in $\tilde{O}(m+{\sf poly}(k)n)$ time.
The arguments in \cite{PY} are complex, and here by a simpler proof we 
obtain the following generalization as well as an improvement on the {\mcut} query.
For a set $S \subs V$ of terminals we say that a graph is {\bf $k$-$S$-connected} if $\ka(s,t) \geq k$ for all $s,t \in S$.
We will improve over Theorem~\ref{t1} for $k$-$S$-connected graphs as follows.

\begin{theorem} \label{t3}
For any $k$-$S$-connected graph with $|S| \geq 3k$,
there exists an $O(k|S|)$ space data structure, that includes a list of  $3|S|$ cuts, 
and answers {\econ} and {\mcut} queries for node pairs in $S$ in $O(1)$ time.
\end{theorem}

Combining the data structures in Theorems \ref{t2} and \ref{t3}, we get the following, see also Table~\ref{tbl:cut}.

\begin{theorem} \label{t4}
There exists an $O(k^2 n)$ space data structure that includes a list of $O(kn)$ cuts, 
that answers {\econ} and {\cut} queries in $O(1)$ time; 
a list of $O(k^2 n)$ cuts and space $O(k^3 n)$ enables to answer also {\mcut} queries in $O(1)$ time.
\end{theorem}

We note that the \cite{PSY} data structure can be augmented by a list of $O(kn \log n)$ cuts to support {\mcut} queries in time $O(\log n)$.
Our query time $O(1)$ is always better, and our cut list size $O(k^2n)$ is better when $k < \log n$.
Furthermore, if we want to answer only {\cut} queries, then our list size is $O(kn)$, 
which is smaller than the cut list size $O(kn \log n)$ of \cite{PSY}.

All our data structures can be constructed in polynomial time;
we will not  discuss designing efficient construction algorithms here and leave this for future work. 

The preliminary version of this paper is \cite{N-ds}. 
We note that Theorems \ref{t1} and \ref{t2} did not appear in the preliminary version \cite{N-ds},
which had only Theorems \ref{t3} and \ref{t4},
and also some minor results, that are now dominated by Theorems \ref{t1} and \ref{t2}. 

Theorems \ref{t2}, \ref{t3}, and \ref{t4} are proves in section \ref{s2}, \ref{s3}, and \ref{s4}, respectively.  

\section{Connectivity queries in general graphs (Theorem~\ref{t2})} \label{s2}

We start by describing a simple data structure for answering edge connectivity queries. 
Let $\la(s,t)=\la_G(s,t)$ denote the maximum number of edge disjoint $st$-paths in $G$. 
For $\ell \leq k$ let us define the relation $\RR^\ell=\{(s,t) \in V \times V: \la(s,t) \geq \ell\}$. 
It is known that $\RR^\ell$ is an equivalence relation and its equivalence classes are called 
{\bf classes of $\ell$-edge-connectivity}, or just {\bf $\ell$-classes} for short. 
Let $\PP^\ell$ denote the partition of $V$ into $\ell$-classes.
One can see that if $(s,t) \in \RR^{\ell+1}$ then $(s,t) \in \RR^\ell$, and this implies that  
$\PP^{\ell+1}$ is a refinement of $\PP^\ell$.
Consequently, the partitions $\PP^0, \ldots, \PP^{k+1}$ form a laminar (multi-)family 
and can be represented by by a leveled rooted tree $T$ as follows:
\begin{itemize}
\item
The nodes of  each level $\ell=0,\ldots,k+1$ are the parts of $\PP^\ell$.
\item
The parent of a part $Q \in \PP^{\ell+1}$ is the part $P \in \PP^\ell$ that contains $Q$.
\end{itemize}
We may add the partition into singletons as level $k+2$, and assume that the leaf set of $T$ is $V$. Note that then
$\min\{\la(s,t),k+1\} =  \ell(\lca_T(s,t))$, where $\ell(v)$ is the the distance of a node $v$ of $T$ to the root. 
Thus augmenting $T$ by Gabow \& Tarjan \cite{GT} lowest common ancestor data structure enables to find $\ell(\lca_T(u,v))$ in $O(1)$ time. 
Note that $T$ has $O(kn)$ edges. 
The size of $T$ can be reduced to $O(n)$ by shortcutting nodes that have a unique child, 
so we will get a leveled tree $\langle T,\ell \rangle$ of size $O(n)$.
And since the parameter $k$ has no role in the size of $T$, we may set $k=\infty$ and obtain
$\la(s,t)=  \ell(\lca_T(s,t))$.

The reader may observe that this data structure is not really related to connectivity, 
as it does not use any special property (e.g., submodularity) of the cut function. 
Let us illustrate this on a more general setting. 
We say that $A \subs V$ is an {\bf $st$-set} if $s \in A$ and $t \notin A$. 
Let $f$ be an arbitrary non-negative integer valued set function on subsets of $V$.
Let us define the {\bf $f$-connectivity} between $s,t$ by 
$\la_f(s,t)=\min\{f(A):A \mbox{ is an } st\mbox{-set or a } ts\mbox{-set}\}$. 
W.l.o.g. we may assume that $f$ is {\bf symmetric}, namely, that $f(A)=f(V \sem A)$ for all $A$,
as otherwise we may consider the function $g(A)=\min\{f(A),f(V\sem A\}$. 
Note that then $\la_f(s,t)=\min\{f(A):A \mbox{ is an } st\mbox{-set}\}$.
By the same method we just described, one can deduce the following.

\begin{corollary}
For any set function $f$ on $V$ there exist
a leveled tree $\langle T,\ell \rangle$ with leaf set $V$ and $O(n)$ edges such that
$\la_f(s,t)=  \ell(\lca_T(s,t))$ for any $s,t \in V$.
\end{corollary}

We emphasize again that the above corollary is valid for {\bf any} set function. 
For edge connectivity, the appropriate set function is given by $f(A)=|\de(A)|$, 
where $\de(A)=\de_G(A)$ is the set of edges in $G$ with exactly one end in $A$. 

While a natural way to represent an edge cut of a graph is by a node subset, 
for node cuts we need a more general object given in the following definition.

\begin{definition}
A {\bf biset} on a groundset $V$ is an ordered pair $\A=(A,A^+)$ such that $A \subs A^+ \subs V$;
$A$ is the {\bf inner part} and $A^+$ is the {\bf outer part} of $\A$, $\p\A=A^+\sem A$ is the {\bf cut} of $\A$,
$A^*=V \sem A^+$ is the {\bf co-set} of $\A$, and $\A^*=(V \sem A^+,V \sem A)$ is the {\bf co-biset} of $\A$.
We say that $\A$ is an {\bf $st$-biset} if $s \in A$ and $t \in A^*$. 
\end{definition}

Let $f$ be an arbitrary biset function on a groundset $V$. 
W.l.o.g. we will assume that $f$ is {\bf symmetric}, namely, that $f(\A)=f(\A^*)$ for all $\A$.
Given (a value oracle for) such $f$ we define the {\bf $f$-connectivity} between $s,t$ by 
$$
\la_f(s,t)=\min\{f(\A):\A \mbox{ is an } st\mbox{-biset}\} \ .
$$

For node connectivity of a given graph $G$, an appropriate biset function $f$ is given by $f(\A)=|\p\A|+|\de(\A)|$, where here $\de(\A)=\de_G(\A)$ 
is the set of edges in $G$ with one end in $A$ and the other in $A^*$. 
We will prove the following generalization of Theorem~\ref{t2}.

\begin{theorem}  \label{t:f}
For any biset function $f$ on $V$ and a positive integer $k$ there exist exists
a leveled tree $\langle T,\ell \rangle$ with $O(n)$ edges and
an edge weighted graph $\langle H=(V,F),w \rangle$ with $O(k\ga n)$ edges, where $\ga=\max\{|\p\A|:f(\A) \leq k\}$, 
such that for any $s,t \in V$ the following holds: 
\begin{equation} \label{e:laf}
\min\{\la_f(s,t),k+1\} = \left \{ \begin{array}{ll}
w(st)                  & \ \mbox{if} \ \ uv \in F \\
 \ell(\lca_T(s,t))  & \mbox{ otherwise}
\end{array} \right .
\end{equation}
\end{theorem}

To see that Theorem~\ref{t:f} implies Theorem~\ref{t2}, note that if $f(\A)=|\p\A|+|\de(\A)|$
then $\ga=\max\{|\p\A|:f(\A) \leq k\}=k$, and thus for this case both theorems coincide. 

In the rest of this section we prove Theorem~\ref{t:f}. 
We may assume that $f$ is {\bf symmetric}, namely, that $f(\A)=f(\A^*)$ for every biset $\A$; 
otherwise, we may consider the function $g(\A)=\min\{f(\A),f(\A^*)\}$.
We will start with a simpler task -- 
given a set family $\FF$ and $S \subs V$, design a data structure 
that determines for $s,t \in S$ whether there exists and $st$-biset $\A \in \FF$.

\begin{definition}
Let $\FF$ be a symmetric biset family on $V$. 
We say that $s,t \in V$ are {\bf $\FF$-inseparable} if $\FF$ has no $st$-biset. 
The {\bf $\FF$-inseparability graph} $H_S$ of $S \subs V$ has node set $S$ 
and edge set $\{st: s,t \in S \mbox{ are } \FF\mbox{-inseparable}\}$.
\end{definition}

One can observe that for any $S' \subs S$, if $H_S$ is the $\FF$-separability graph of $S$ 
then $H_S[S']$ is the $\FF$-separability graph of $S'$.
The following lemma was proved by Hsu \& Lu in \cite{HL} for the particular case when 
$\FF=\{\A: \de_G(\A)=\empt, |\p\A| \geq k+1\}$ for a given a graph $G$;
the proof for an arbitrary (symmetric) biset family is similar. 

\begin{lemma} \label{l:partition}
Let $\FF$ be a symmetric biset family on $V$ and let $H_S$ be the $\FF$-inseparability graph of $S \subs V$.
There exists an ordered partition $\PP=(S_1, \ldots,S_q)$ of $S$ such that in $H_S$
each $S_i$ is a clique with at most $\max\{2\ga-1,0\}$ neighbors in $S_{i+1} \cup \cdots \cup S_q$, 
where $\ga=\max\{|\p\A \cap S| :A \in \FF\}$. 
Consequently, if $F_S$ is the set of edges in $H_S$ that do not belong to any clique $S_i$ 
then $|F_S| \leq |S|\max\{2\ga-1,0\}$.
\end{lemma}
\begin{proof}
We first show that $H_S$ has a clique $C \subs V$ with at most $\max\{2\ga-1,0\}$ neighbors. 
The algorithm for finding such a clique $C$ is as follows.

\medskip

\begin{algorithm}[H]
\caption{{\bf Find-Clique$(S,H_S,\FF)$}}
\label{alg:C}
$R \gets \empt$ \\
\While{\em $S \sem R$ is not a clique in $H$}
{
let $s,t \in S \sem R$ be two non-adjacent nodes in $H_S$ \\
let $\A \in \FF$ be an $st$-biset or a $ts$-biset with $|A \cap S \cap R| \leq |A^* \cap S \cap R|$ \\
$R \gets R \cup \p\A$, $S \gets S \sem A^*$, $H_S \gets H \sem A^*$ 
}
\Return{$C=S \sem R$}  
\end{algorithm}

\medskip

We show that the algorithm is well defined. 
\begin{itemize}
\item
The condition of the while-loop implies the existence of $s$ and $t$ in step 3 and 
also of $\A$ in step 4 (since $\FF$ is symmetric). 
\item
Step 5 removes some node from $S \sem R$ but also keeps some node in $S \sem R$. 
Hence the while-loop terminates in at most $|S|$ iterations and $|C| \geq 1$ at the end of the algorithm. 
\end{itemize}
It remains to show that $C$ has at most $\max\{2\ga-1,0\}$ neighbors. 
Let $r_0=0$ and $r_i=|R \cap S|$ at the end of iteration $i$.
By steps 4,5,6 in the algorithm, $r_i \leq |\p\A_i|+\f{1}{2}r_{i-1} \leq \ga+\f{1}{2}r_{i-1}$,
where $\A_i$ is the biset chosen at iteration $i$.
Now we continue by induction. Clearly, $r_0=0 \leq \max\{2\ga-1,0\}$.
For the induction step assume that $r_{i-1} \leq  \max\{2\ga-1,0\}$. Then 
$$
r_i \leq |\p\A_i \cap S|+\f{1}{2}r_{i-1} \leq \ga+\f{1}{2} \max\{2\ga-1,0\}=\max\{2\ga-1/2,\ga\} \ . 
$$
Since $r_i,\ga$ are integers we get $r_i \leq  \max\{2\ga-1,\ga\}=\max\{2\ga-1,0\}$, as claimed. 

Now we show that $S$ has an ordered partition $\PP$ as in the lemma.
The algorithm for obtaining such $\PP$ is as follows. 

\medskip

\begin{algorithm}[H]
\caption{{\bf Clique-Partition$(S,H_S,\FF)$}}
\label{alg:CP}
$q \gets 1$ \\
\While{\em $S \neq \empt$}
{
$S_q \gets  \mbox{Find-Clique}(S,H_S,\FF)$ \\
$S \gets S \sem S_q$, $H_S \gets H_S \sem S_q$ \\
$q \gets q+1$ \\
}
\Return{$\PP=(S_1, \ldots, S_q)$}  
\end{algorithm}

\medskip

Since at any step of Algorithm~\ref{alg:CP}, $S_q$ has at most $\max\{2\ga-1,0\}$ neighbors in $S \sem S_q$, 
the lemma follows. 
\end{proof}

We now finish the proof of Theorem~\ref{t:f}. 
Let $\FF^\ell=\{\A:f(\A) \leq \ell\}$ and let $H^\ell$ be the $\FF^{\ell}$-inseparability graph of $V$ , $\ell=1, \ldots, k+1$. 
Note that $\FF^{\ell-1} \subs \FF^\ell$ and thus $H^{\ell-1} \supseteq H^\ell$. 
The algorithm starts with the $0$-level ordered partition $\PP^0 \gets \{V\}$ and continues with iterations. 
At the beginning of iteration $\ell \geq 1$ we already have an $(\ell-1)$-level ordered partition $\PP^{\ell-1}=(S_1, \ldots, S_q)$ of $V$.
For each part $S_i$ of $\PP^{\ell-1}$ we compute 
an ordered partition $\PP_i$ of $S_i$ and an edge subset $F_i$ of $H^\ell$
as in Lemma~\ref{l:partition} with $S=S_i$ and $\FF=\FF^\ell$.
We then concatenate the computed partitions into one ordered partition $\PP^\ell=(\PP_1, \ldots, \PP_q)$
and let $F^\ell=\cup_{i=1}^q F_i$.
Note that $F_i \leq |S_i|\max\{2\ga-1,0\}$, hence $|F^\ell| \leq \sum_{i=1}^q| S_i|\max\{2\ga-1,0\}=n\max\{2\ga-1,0\}$. 
For each $st \in F^\ell$ we let $w(st)=\min\{\la_f(u,v),k+1\}$; note that  $w(s,t) \geq \ell$ and that an inequality may hold. 

Note that $\PP^\ell$ is a refinement of $\PP^{\ell-1}$.
We augment the partitions $\PP^0, \ldots \PP^{k+1}$ by the partition into singletons, 
and represent the obtained sequence of partitions by 
a leveled tree $\langle T,\ell \rangle$ of size $O(n)$, as was described at the beginning of this section
for the edge connectivity case. 
In addition, we let $F=\cup_{\ell=1}^{k+1}F^\ell$, so $|F| \leq (k+1)n\max\{2\ga-1,0\}=O(kn \ga)$.

It remains to show that (\ref{e:laf}) holds for $\langle T,\ell \rangle$ and $\langle H=(V,F),w \rangle$.
Consider a pair $s,t \in V$.
If $st \in F$ then (\ref{e:laf}) holds by the definition of $w$.
Else, if $st \notin F$ then $s,t$ belong to the same part of $\PP^\ell$ for every $\ell \leq \la_f(s,t)$,
and belong to distinct parts of $\PP^{\ell+1}$; this implies that $\min\{\la_f(u,v),k+1\} = \ell(\lca(u,v))$.

This concludes the proof of Theorem~\ref{t:f}, and thus also the proof of Theorems \ref{t2} and \ref{t1} is also complete. 

\section{Connectivity and cut queries in {\em k-S}-connected graphs (Theorem \ref{t3})} \label{s3}

We start by giving a simple proof for a slight improvement of Theorem~\ref{t3} for the case $S=V$.
This case was considered by Pettie-Yin in \cite{PY}, 
but our data structure and arguments are substantially simpler than those in \cite{PY}. 
Here, and in other parts of the paper we will need the following known lemma, 
for which we provide a proof for completeness of exposition.

\begin{lemma} \label{l:color}
Let $H$ be a directed graph of maximum indegree $d$. 
Then the underlying graph of $H$ is $(2k+1)$-colorable, and such a coloring can be computed in linear time. 
\end{lemma}
\begin{proof}
Since the indegree of every node in $H$ is at most $k$,
every subgraph of the underlying graph of $H$ has a node of degree $2k$.
A graph is {\bf $d$-degenerate} if every subgraph of it has a node of degree $d$. 
It is known that any $d$-degenerate graph can be colored with $d+1$ colors, in linear time, see \cite{EH,MB}. 
This implies that the underlying graph of $H$ can be colored with $2k+1$ colors in linear time. 
\end{proof}

\begin{lemma} \label{l:V}
For any $k$-connected graph $G=(V,E)$
there exists an $O(kn)$ space data structure with a list of  $2n$ cuts 
that answers {\econ} and {\mcut} queries in $O(1)$ time.
\end{lemma}
\begin{proof}
Let $K$ be the set of nodes of degree $k$ in $G$.
If $s \in K$ then $\de_G(s)$ is a minimum $st$-cut for all $t$. There are $|K|$ such minimum cuts.
This situation can be recognized in $O(1)$ time, hence assume that $s,t \in V \sem K$.
Let $F$ be the set of edges $st \in E$ such that $s,t \in V \sem K$ and $\ka_G(s,t)=k$.
By Mader's Critical Cycle Theorem \cite{M}, $F$ is a forest, so $|F| \leq n-|K|-1$.
Thus just specifying the nodes in $K$ and edges in $F$ and a list of $|F| +|K| \leq n$ minimum $st$-cuts 
gives an $O(kn)$ space data structure that answers the relevant queries in $O(1)$ time. 

Henceforth assume that $s,t \in V\sem K$ and $st \notin E$.
We will show that then there exists a list of $n-|K|$ cuts, such that whenever $\ka(s,t)=k$
there exists a minimum $st$-cut in the list.
We will also show how to choose the right minimum cut from the list in $O(1)$ time.

Let $A^*=V \sem (A \cup \p\A)$ denote the ``node complement'' of $A$. 
We say that $A$ is: 
a {\bf tight set} if $A,A^* \neq \empt$ and $|\p A|=k$, 
an {\bf $st$-set} if $s \in A$ and $t \in A^*$,  and
a {\bf small set} if $|A| \leq \f{n-k}{2}$. 
Note that $A$ is tight if and only if $\p A$ is a minimum cut of $G$, 
and $A$ is a union of some, but not all, connected components of $G \sem \p A$.
The following statement is a folklore, c.f. \cite{N-subs,N-impr}.

\begin{claim} \label{c:tight}
{\em Let $A,B$ be tight sets. 
If the sets $A \cap B^*,B \cap A^*$ are both nonempty then $A,B$ are both tight.
If $A,B$ are small sets and $A \cap B \neq \empt$ then $A \cap B$ is tight.}
\end{claim}

Let $R=\{s \in V \sem K:\mbox{there exist a small tight set containing } s\}$.
For $s \in R$ let $C_s$ be the (unique, by Claim~\ref{c:tight}) inclusion minimal small tight set that contains $s$. 
Let $\CC=\{C_s:s \in R\}$. 
We claim that the family $\{\p C:C \in \CC\}$ is a ''short'' list of $n-|K|$ minimum cuts, 
that for every $s,t \in V \sem K$ with $st \notin E$ includes a minimum $st$-cut.
Specifically, we claim that $\ka(s,t)=k$ if an only if at least one of the following holds: 
\begin{enumerate}[(i)]
\item
$s \in R$ and $C_s$ is an $st$-set, or 
\item
$t \in R$ and $C_t$ is a $ts$-set.
\end{enumerate}
Indeed, if (i) holds then $\p C_s$ is a minimum $st$-cut, while if (ii) holds then $\p C_t$ is a minimum $st$-cut. 
Thus $\ka(s,t)=k$ if (i) or (ii) holds. 
Assume now that $\ka(s,t)=k$ and we will show that (i) or (ii) holds.
Let $Q \subset V \sem \{s,t\}$ be a minimum $st$-cut. 
Then one component $A$ of $G \sem Q$ contains $s$ and the other $B$ contains $t$.  
Since $|A|+|B| \leq n-|Q| =n-k$, one of $A,B$, say $A$ is small. Thus $s \in R$. 
Since $C_s \subs A$ and since $t \in A^*$, we have $t \in C_s^*$. 
Consequently, $\p C_s$ is a minimum $st$-cut, as required.

Finally, we will show that $\CC$ can be partitioned into at most $2k+1$ laminar families.  
Consider two sets $A=C_a$ and $B=C_b$ that are not laminar. 
Then $a \notin A \cap B$ since otherwise by Claim~\ref{c:tight} $A \cap B$ is a (small) tight set that contains $a$,
contradicting the minimality of $A=C_a$.
By a similar argument, $b \notin A \cap B$. 
We also cannot have both $a \in A \cap B^*$ and $b \in B \cap A^*$, as then by Claim~\ref{c:tight} $A \cap B^*$ is a tight set that contains $a$,
contradicting the minimality of $A=C_a$.
Consequently, $a \in \p B$ or $b \in \p A$. 
Construct an auxiliary directed graph $H$ on node set $R$ and edges set $\{ab:a \in \p C_b\}$.
The indegree of every node in $H$ is at most $k$.
By Lemma~\ref{l:color}, we can compute in polynomial time a partition of $R$ into at most $2k+1$ independent sets. 
For each independent set $R_i$, the family $\{C_s:s \in R_i\}$ is laminar. 

Our data structure for pairs $s,t \in V\sem K$ with $st \notin E$ consists of:
\begin{itemize}
\item
A family $\TT$ of at most $2k+1$ trees, 
where each tree $T \in \TT$ with a mapping $\psi_T: V \rightarrow V(T)$
represents one of the at most $2k+1$ laminar families of tight sets as;
the total number of edges in all trees in $\TT$ is at most $n-|K|$.
\item
For each tree $T \in \TT$, a linear space data structure that answers ancestor/descendant queries in $O(1)$ time.
This can be done by assigning to each node of $T$ the in-time and the out-time in a DFS search on $T$.
\item
A list $\{\p C_s:s \in R\}$ of $|R|=n-|K|$ minimum cuts;
this can be also encoded by an auxiliary directed graph $H=(V,F)$ with edge set $F=\{ts:t \in \p C_s\}$.  
Using perfect hashing data structure we can check whether $ts \in F$ in $O(1)$ time.
\end{itemize}

For every $s \in S$ let $T_s$ be the (unique) tree in $\TT$ where $C_s$ is represented.
A direct consequence of (i) and (ii) above specifies how we answer the queries.
\begin{enumerate}[(i)]
\item
If in $T_s$, $\psi_{T_s}(t)$ is not a descendant of $\psi_{T_s}(s)$ and $t \notin \p C_s$, then $\p C_s$ is a minimum~$st$-cut.
\item
If in $T_t$, $\psi_{T_t}(s)$ is not a descendant of $\psi_{T_t}(t)$ and $s \notin \p C_t$, then $\p C_t$ is a minimum~$st$-cut.
\end{enumerate}
If none of  (i),(ii) holds then $\ka(s,t) \geq k+1$.

It is easy to see that with appropriate pointers, and using perfect hashing data structure 
to check adjacency in the auxiliary directed graph $H$, 
we get an $O(kn)$ space data structure that checks the conditions above in $O(1)$ time. 
If one of the conditions holds, the data structure return a pointer to one of $\p C_s$ or $\p C_t$. 
Else, it reports that $\ka(s,t) \geq k+1$. 

This concludes the proof of the lemma.
\end{proof}
 
In the rest of this section let $G=(V,E)$ be a $k$-$S$-connected graph with $|S| \geq 3k$. 
Our proof of Theorem~\ref{t3} uses ideas similar to those in the proof of Lemma~\ref{l:V}, 
but it is substantially more involved.
One reason is that we cannot use Mader's Critical Cycle Theorem \cite{M}.
Another reason is that for the simpler case $S=V$ we could relate cuts to node subsets,  
but for the more general subset $k$-connectivity case, we again need to consider bisets.
Here we will say that $\A$ is a {\bf tight biset} if  $A \cap S \neq \empt$, $A^* \cap S \neq \empt$, and 
$|\p \A|+|\delta(\A)|=k$, where $\delta(\A)$ is the set of edges in $G$ that go from $A$ to $A^*$.
Note that $\A$ is tight if and only if $\p \A \cup \delta(\A)$ is a minimum $st$-cut for some $s \in A \cap S$ and $t \in A^* \cap S$.
We will consider the family $\FF=\{(A \cap S, A^+ \cap S):\A \mbox{ is tight}\}$ 
obtained by projecting the tight bisets on $S$.
Note that for $\A \in \FF$ there might be many tight bisets in $G$ whose projection on $S$ is $\A$,
and that there always exists at least one such biset. 

\begin{definition} 
The {\bf intersection} and the {\bf union} of two bisets $\A,\B$ are the bisets defined by
$\A \cap \B = (A \cap B, A^+ \cap B^+)$ and $\A \cup \B=(A \cup B,A^+ \cup B^+)$.
The biset $\A \sem \B$ is defined by $\A \sem \B=(A \sem B^+,A^+ \sem B)$.
We say that $\A,\B$: 
{\bf intersect} if $A \cap B \neq \empt$,
{\bf cross} if $A \cap B \neq \empt$ and $A^* \cap B^* \neq \empt$, and 
{\bf co-cross} if $A \cap B^* \neq \empt$ and $B \cap A^* \neq \empt$.
\end{definition}

We say that $\A \in \FF$ is a {\bf small biset} if $|A| \leq \f{|S|-k}{2}$, and $\A$ is a {\bf large biset} otherwise.  
Clearly, $|\p\A| \leq k$ for all $\A \in \FF$. The family $\FF$ has the following properties, c.f. \cite{N-subs,N-impr}.

\begin{lemma} \label{l:prop}
The family $\FF=\{(A \cap S, A^+ \cap S):\A \mbox{ is tight}\}$ has the following properties:
\begin{enumerate}
\item
$\FF$ is {\bf symmetric}: $\A^* \in \FF$ whenever $\A \in \FF$.
\item
$\FF$ is {\bf crossing}: $\A \cap \B, \A \cup \B \in \FF$ whenever $\A,\B \in \FF$ cross.
\item
$\FF$ is {\bf co-crossing}: $\A \sem \B,\B \sem \A \in \FF$ whenever $\A,\B \in \FF$ co-cross. 
\item
If $\A,\B \in \FF$ are small intersecting bisets then $\A \cap \B,\A \cup \B \in \FF$.
\end{enumerate}
\end{lemma}

\begin{definition}
We say that a biset {\bf $\B$ contains a biset $\A$} and write $\A \subs \B$ 
if $A \subs B$ and $A^+ \subs B^+$. 
$\A,\B$ are {\bf laminar} if one of them contains the other or if $A \cap B=\empt$.
A biset family is laminar if its members are pairwise laminar.
\end{definition}

For every $s \in S$ let $\CC_s$ be the family of all inclusion minimal bisets $\C \in \FF$ with $s \in C$ and $|C| \leq |C^*|$.
Let $\CC=\cup_{s \in S}\CC_s$.
Note that $|\CC_s| \leq |S|-1$ and that $\CC_s$ can be computed using $|S|-1$ min-cut computations.  
The following observation follows from the symmetry of $\FF$.

\begin{lemma}
Let $\A \in \FF$ be an $st$-biset. If $|A| \leq |A^*|$ then there is an $st$-biset in $\CC_s$,
and if  $|A^*| \leq |A|$ then there is a $ts$-biset in $\CC_t$. 
Consequently, for any $s,t \in S$, the family $\CC$ contains an $st$-biset or a $ts$-biset.
\end{lemma}

The next lemma shows that $|\CC_s| \leq 3$ if $|S| \geq 3k$.

\begin{lemma} \label{l:Cs}
For any $s \in S$, $\CC_s$ contains at most one small biset and at most $\f{2(|S|-1)}{|S|-k}$ large bisets.
In particular, if $|S| \geq 3k$ then in $\CC_s$ there are at most two large bisets and $|\CC| \leq 3|S|$.
\end{lemma}
\begin{proof}
In $\CC_s$ there is at most one small biset, by property~4 in Lemma~\ref{l:prop}. 
No two large bisets $\A,\B \in \CC_s$ cross, as otherwise $s \in \A \cap \B$ 
and (by property~2 in Lemma~\ref{l:prop}) $\A \cap \B \in \FF$, 
contradicting the minimality of $\A,\B$.
Thus the sets in $\{C^*:\C \in \CC_s\}$ are pairwise disjoint.
Furthermore, $|C^*| \geq |C| \geq \f{|S|-k}{2}$ for every large biset $\C \in \CC_s$.
This implies that the number of large bisets in $\CC_s$ is at most 
$\f{|S|-1}{(|S|-k)/2} \leq \f{2(3k-1)}{2k} <3$ if $|S| \geq 3k$.
\end{proof}

By a proof similar to the one of Lemma~\ref{l:V} we have the following.

\begin{lemma} \label{l:S-lam}
The family of small bisets in $\CC$ can be partitioned in polynomial time into at most $2k+1$ laminar families.
\end{lemma}  
\begin{proof}
Consider two small bisets $\A \in \CC_a$ and $\B \in \CC_b$ that are not laminar. 
Then $a \notin A \cap B$ since otherwise $\A \cap \B \in \FF$, 
contradicting the minimality of $\A$. Similarly, $b \notin A \cap B$. 
We also cannot have both $a \in A \cap B^*$ and $b \in B \cap A^*$, as then 
$\A,\B$ co-cross and thus $\A \sem \B,\B \sem \A \in \FF$, 
contradicting the minimality of $\A,\B$.
Consequently, $a \in \p B$ or $b \in \p A$. 
Construct an auxiliary directed graph $H$ on node set $R$ and edges set $\{ab:a \in \p C_b\}$.
The indegree of every node in $H$ is at most $k$.
By Lemma~\ref{l:color}, we can compute in polynomial time a partition of $R$ into at most $2k+1$ independent sets. 
For each independent set $R_i$, the family $\{C_s:s \in R_i\}$ is laminar. 
\end{proof}

Later, we will prove the following.

\begin{lemma} \label{l:S-lam'}
If $|S| \geq 3k$ then the family of large bisets in $\CC$ can be partitioned in polynomial time into at most $3(2k+1)$ laminar families.
\end{lemma}  

Lemmas \ref{l:S-lam} and \ref{l:S-lam'} imply that $\CC$ can be partitioned into at most $4(2k+1)$ laminar families. 
For our purposes, we just need the family of the inner sets of each family to be laminar.  
Together with Lemma~\ref{l:Cs} this implies Theorem~\ref{t2} by the same way as in the proof of Lemma~\ref{l:V},
except the following minor differences.
\begin{itemize}
\item
Here we have $4(2k+1)$ laminar families instead of $2k+1$ laminar families, and 
for each $s \in S$ we have by Lemma~\ref{l:Cs} $|\CC_s| \leq 3$ instead of $|\CC_s| = 1$.
\item
For each biset $\C \in \CC$, our list of cuts will include a mixed cut $\p \A \cup \delta(\A)$ of some tight biset of $G$
whose projection on $S$ is $\C$.
\end{itemize}

These differences affect space and query time only by a small constant factor.
Thus all we need is to prove Lemma~\ref{l:S-lam'}, which we will do in the rest of this section.

\begin{figure}
\centering 
\includegraphics{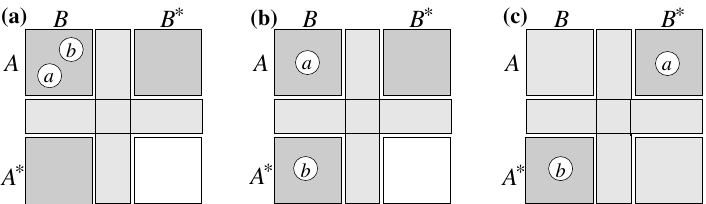}
\caption{Illustration to the proof of Lemma \ref{l:lami}; dark gray sets are non-empty.}
\label{f:lami}
\end{figure}

\begin{lemma} \label{l:lami}
Let $\A \in \CC_a$ and $\B \in \CC_b$ be two non-laminar large bisets in $\CC$. 
If $|S| \geq 3k$ then $a \in \p \B$ or $b \in \p\A$, or (see Fig.~\ref{f:lami}(a)):
$a,b \in A \cap B$, $A^* \cap B^*=\empt$, and both $A \cap B^*,B \cap A^*$ are non-empty.
\end{lemma}
\begin{proof}
Assume that $a \notin \p\B$ and $b \notin \p\A$. We will show that then the case in Fig.~\ref{f:lami}(a) holds.

Suppose that $a \in A \cap B$; the analysis of the case $b \in A \cap B$ is similar. 
Then $A^* \cap B^*=\empt$;
otherwise, $\A,\B$ cross and (by property~2 in Lemma~\ref{l:prop})
we get $\A \cap \B \in \FF$, contradicting the minimality of $\A$.  
Furthermore, if $B \cap A^*=\empt$ we get 
$\f{|S|-k}{2} < |A| \leq |A^*| = |\p\B \cap A^*| \leq k$, contradicting that $|S| \geq 3k$.
By a similar argument, $A \cap B^* \neq \empt$. 
If $a \in A \cap B$ and $b \in B \cap A^*$ (Fig.~\ref{f:lami}(b)), then $\A,\B$ co-cross 
(by property~3 in Lemma~\ref{l:prop}) and thus $\B \sem \A \in \FF$;
this contradicts the minimality of $\B$. 

If none of $a,b$ is in $A \cap B$, then (see Fig.~\ref{f:lami}(c)) $a \in A \cap B^*$ and $b \in B \cap A^*$. 
Consequently, $\A,\B$-co-cross, and thus (by property~3 in Lemma~\ref{l:prop}) 
$\A \sem \B,\B \sem \A \in \FF$, contradicting the minimality of $\A,\B$. 

Thus the only possible case is the one in Fig.~\ref{f:lami}(a), completing the proof of the lemma.
\end{proof}

From Lemma~\ref{l:lami}, by a proof identical to that as in Lemma~\ref{l:V} we get the following.

\begin{corollary} \label{c:lami}
The family of large bisets in $\CC$ can be partitioned in polynomial time 
into at most $2k+1$ parts such that 
any two bisets $\A \in \CC_a$ and $\B \in \CC_b$ that belong to the same part $\PP$
are either laminar, or have the following property:
$a,b \in A \cap B$ and $A^* \cap B^*=\empt$. 
\end{corollary}

Thus the following lemma finishes the proof of Lemma~\ref{l:S-lam'}, and also of Theorem~\ref{t3}.

\begin{lemma} \label{l:bipartite}
Let $\PP$ be one of the $2k+1$ parts as in Corollary~\ref{c:lami};
in particular, if $\A,\B \in \PP$ are not laminar then  $A \cap B \neq \empt$ and $A^* \cap B^*=\empt$.
Then $\PP$ can be partitioned in polynomial time into at most $3$ laminar families. 
\end{lemma}
\begin{proof}
Let ${\cal M}$ be the family of maximal members in $\PP$. We will show later that 
the set family $\HH=\{A^*:\A \in {\cal M}\}$ has a hitting set $U$ of size $|U| \leq 3$. Now note that:
\begin{itemize}
\item
For every $v \in S$ the family $\PP_v=\{\A \in \PP: v \in A^*\}$ is laminar, since 
$v \in A^* \cap B^*$ for any $\A,\B \in \PP_v$, 
while $A^* \cap B^*=\empt$ for any non-laminar $\A,\B \in \PP$.
\item
Since $U$ is a hitting set of $\HH$, for any $\A \in \PP$ there is $v \in U \cap A^*$, and then $\A \in \PP_v$.
\end{itemize}
Summarizing, each one of the families $\PP_v$ is laminar and $\cup_{v \in U} \PP_v=\PP$.
By removing bisets that appear more than once, we get a partition of $\PP$ into $|U| \leq 3$ laminar families.  

It remains to show that $\HH$ has a hitting set of size $\leq 3$.
A {\bf fractional hitting set} of $\HH$ is a function $h:S \rightarrow [0,1]$ such that
$h(A) = \sum_{v \in A} h(v) \geq 1$ for all $A \in \HH$.
For $v \in S$ let $\HH_v$ be the family of sets in $\HH$ that contain $v$, and let ${\cal M}_v=\{\A \in {\cal M}:A^* \in \HH_v\}$. 
Note that:
\begin{itemize}
\item
$h(v)=\f{2}{|S|-k+1}$ for all $v \in S$ is a fractional hitting set of $\HH$ and $h(S)=\f{2|S|}{|S|-k+1} < 3$.
\item
No two bisets in ${\cal M}_v$ intersect and $|\HH_v| \leq |{\cal M}_v|$.
This implies $|\HH_v| \leq |{\cal M}_v| \leq \f{2|S|}{|S|-k+1} < 3$, so $|\HH_v| \leq 2$ for all~$v \in S$. 
\end{itemize}
Since $|\HH_v| \leq 2$ for all $v \in S$, computing a minimum hitting set of $\HH$ 
reduces to the minimum Edge-Cover problem, and $\HH$ has a hitting set $U$ of size $|U| \leq \f{4}{3} \cdot h(S)$
($\f{4}{3}$ is the integrality gap of the Edge-Cover problem).
Since $\f{4}{3} \cdot h(S) < \f{4}{3} \cdot 3$, $\HH$ has a hitting set $U$ of size $|U| \leq 3$.
\end{proof}

This concludes the proof of Lemma~\ref{l:S-lam'} and thus also the proof of Theorem \ref{t3} is complete. 

\section{Cut queries in general graphs (Theorem \ref{t4})} \label{s4}

In the first part of Theorem~\ref{t4} we need to design  an $O(k^2 n)$ space data structure with a list of $O(kn)$ cuts, 
that answers {\econ} and {\cut} queries in $O(1)$ time.
We can use our $O(k^2 n)$-space data structure from Theorem~\ref{t1} to answer {\econ} queries. 
To answer {\cut} queries, we combine our data structure for $k$-$S$-connectivity in Theorem~\ref{t3}
with the Hsu-Lu \cite{HL} data structure. Recall that the \cite{HL} data structure consists of 
an auxiliary graph $H=(V,F)$ with $|F|=O(kn)$ edges  and 
an ordered partition $\PP=(S_1,\ldots, S_q)$ of $V$, 
such that $\ka(s,t) \geq k+1$ iff $s,t$ belong to the same part of $\PP$ or $st \in F$. 
Overall, the \cite{HL} data structure can be implemented using $O(kn)$ space
and answers {\con} queries in $O(1)$ time.

We augment the Hsu \& Lu \cite{HL} data structure by adding to each part $S \in \PP$ 
a data structure for subset $k$-$S$-connectivity;
we add Theorem~\ref{t3} data structure if $|S| \geq 3k$, 
and the trivial data structure (an $S \times S$ matrix) if $|S|<3k$.
Then, by Theorem~\ref{t3} for each part $|S|$ we have the following: 
\begin{itemize}
\item
If $|S| \geq 3k$ then the cut list size is $O(|S|)$ and the other parts use space $O(k|S|)$. 
\item
If  $|S| <3k$ then the cut list size is $O(|S|^2)=O(k|S|)$ and the other parts use space $O(|S|^2)=O(k|S|)$.  
\end{itemize}
The total size of the cut list is bounded by $|F|+k\sum_{S \in \PP} |S|=O(k n)$, and thus uses $O(k^2 n)$ space. 
The size of the other parts is $(kn)$. Thus the total size is $O(k^2 n)$ due to the space required to store the cut list. 
Overall, we use $O(k^2n)$ space and $O(kn)$ cut list size, as required.

In the second part of Theorem~\ref{t4} we need to show  that increasing the cut list size to $O(k^2 n)$ ans space to $O(k^3 n)$ 
enables also to answer {\mcut} queries in $O(1)$ time. For that, we can combine our data structures in Theorems \ref{t1} and \ref{t3}. 
Instead of one clique partition, the Theorem \ref{t1} data structure has $k$ clique partitions, and in addition, $|F|=O(k^2 n)$. 
Thus compared to the first part, the cut list size increases by a factor of $k$ and so is the total space. 
This gives the second part of Theorem \ref{t4}, and thus the proof of Theorem~\ref{t4} is complete. 

\medskip

We can slightly improve the bound on the cut list sizes by a direct short proof, as follows. 

\begin{lemma} \label{l:list}
There exists a list of at most $(2k+1)n$ cuts that contains an $st$-cut 
of size $\leq k$ for any $s,t \in V$ with $\ka(s,t) \leq k$.
Consequently, there exists a list of at most $k(k+2)n$ cuts that contains a minimum $st$-cut 
for any $s,t \in V$ with $\ka(s,t) \leq k$.
\end{lemma}
\begin{proof}
The second part of the lemma easily follows from the first part.
Note that the union of list as in the first part for every $k'=1, \ldots k$ is a list that 
includes a minimum $st$-cut for any $s,t \in V$ with $\ka(s,t) \leq k$. 
The size of this list  is $\sum_{k'=1}^k(2k'+1)n=k(k+2)n$.

We now prove the first part of the lemma. For a biset $\A$ let $\psi(\A)=|\p \A|+|\delta(\A)|$.
Here we will say that $\A$ is an {\bf $st$-tight biset} if $\A$ is an $st$-biset and $\psi(\A)=\ka(s,t)$.
Note that then $\p \A \cup \delta(\A)$ is a minimum $st$-cut, 
and that for any minimum $st$-cut $Q$ there exists an $st$-tight biset $\A$ with $\p \A \cup \delta(\A)=Q$.
It is known that the function $\psi$ satisfies the submodular inequality
$\psi(\A)+\psi(\B) \geq \psi(\A \cap \B)+\psi(\A \cup \B)$, and (by symmetry) also the 
co-submodular inequality $\psi(\A)+\psi(\B) \geq \psi(\A \sem \B)+\psi(\B \sem \A)$.

\begin{figure}
\centering 
\includegraphics{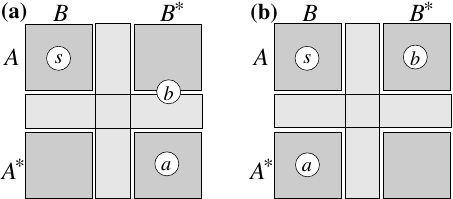}
\caption{Illustration to the proof of Lemma~\ref{l:list}.}
\label{f:s-uncr}
\end{figure}

It is known that if $\A,\B$ are both $st$-tight then so are $\A \cap \B,\A \cup \B$.
Let $\C_{st}$ denote the (unique) inclusion minimal $st$-tight biset.
For $s \in S$ let $T_s=\{t \in V:\ka(s,t) \leq k, |C_{st}| \leq |C_{ts}|\}$.
Let $\CC_s$ be the family of all inclusion minimal bisets in the family $\{\C_{st}:t \in T_s\}$. 
Let $\CC=\cup_{s \in S}\CC_s$. 
One can verify that for any $s,t \in V$ with $\ka_(s,t) \leq k$, 
$\CC$ contains and $st$-biset or a $ts$-biset  $\C$ with $\psi(\C) \leq k$. 
We will show that $|\CC_s| \leq 2k+1$ for all $s \in V$.

Consider distinct bisets $\A=\C_{sa}$ and $\B=\C_{sb}$ in $\CC_s$
We claim that then $a \in \p B$ or $b \in \p A$.
Suppose to the contrary that $a \notin \p B$ and $b \notin \p A$.
If one of $a,b$ is in $A^* \cap B^*$, say $a \in A^* \cap B^*$ (see Fig.~\ref{f:s-uncr}(a)), 
then $\A \cup \B$ is an $sa$-biset and $\A \cap \B$ is an $sb$-biset.Thus 
$$
\ka(s,a)+\ka(s,b)=\psi(\A)+\psi(\B) \geq \psi(\A \cap \B) + \psi(\A \cup \B) \geq \ka(s,b)+\ka(s,a) \ .
$$
Hence equality holds everywhere, so $\A \cap \B$ is $sb$-tight. This contradicts the minimality of $\B$.

Else, $a \in A^* \cap B$ and $b \in B^* \cap A$ (see Fig.~\ref{f:s-uncr}(b)). 
Then $\A \sem \B$ is a $bs$-biset and $\B \sem \A$ is an $as$-biset.Thus 
$$
\ka(s,a)+\ka(s,b)=\psi(\A)+\psi(\B) \geq \psi(\A \sem \B) + \psi(\B \sem \A) \geq \ka(b,s)+\ka(a,s) \ .
$$
Hence equality holds everywhere, so $\A \sem \B$ is $as$-tight and $\B \sem \A$ is $bs$-tight. 
This implies $|C_{sa}| > |C_{bs}|$ and $|C_{sb}| > |C_{as}|$,
and we get the contradiction $|C_{sa}| + |C_{sb}| > |C_{bs}| + |C_{as}|$.

From this we get that $|\CC_s| \leq 2k+1$ for all $s \in V$. To see this, 
construct an auxiliary directed graph $H$ on node set $T_s$ and edge set $\{ab: a \in \p \C_{sb}\}$. 
Note that if $H$ has no edge between $a$ and $b$ then $\C_{sa}=\C_{sb}$. 
The indegree of every node in $H$ is at most $k$.
Thus by Lemma~\ref{l:color} we get that the underlying graph of $H$ is $(2k+1)$-colorable,
and thus $T_s$ can be partitioned into at most $2k+1$ independent sets. 
For each independent set $T'$, the family $\{\C_{st}:t \in T'\}$ consists of a single biset. 

This concludes the proof of the lemma.
\end{proof}

\section{Decomposition of node connectivity into element connectivity} \label{s:elc}

Recall that given a set $S \subs V$ of terminals, 
the element connectivity between $s,t \in S$ is the maximum number of 
pairwise element disjoint $st$-paths, where elements are the edges and the nodes in $V \sem S$.
Let $\ka^S_G(s,t)$ denote the $st$-element connectivity in $G$. 
By Menger's Theorem, $\ka^S_G(s,t)$ equals the minimum size $|C|$ of  a set 
$C$ of elements with $C \cap S=\empt$ such $G \sem C$ has not $st$-path. 
It is easy to see that $\ka_G^S(s,t) \geq \ka_G(s,t)$,
and that an equality holds iff there exists a minimum $st$-cut $C$ with $C \cap S =\empt$. 
We thus will consider the following problem: given a family $\CC$ of subsets of $V$, 
find a ``small'' family $\Sa$ of subsets of $V$ such that for every $s,t \in V$ and $C \in \CC$ with $s,t \notin C$, 
there is $S \in {\cal S}$ with $s,t \in S$ and $C \cap S=\empt$; following \cite{CK}, we will call such a family {\bf $\CC$-resilient}.
The objective can be also to minimize $\sum_{S \in \Sa}|S|$. 
Chuzhoy and Khanna \cite{CK} showed that if $\{C \subset V: |C| \leq k\}$ is the family of all subsets of size $\leq k$,
then there exists a $\CC$-resilient family $\Sa$ of size $|\Sa|=O(k^3 \ln n)$. 
They also gave a randomized polynomial time algorithm for finding such $\Sa$. 
The number of subsets of size $\leq k$ is $\approx n^k$, while 
the relevant family in our case -- as in Lemma~\ref{l:list}, has a much smaller size $|\CC| \leq k(k+2)n$. 
We will consider the case of an arbitrary family $\CC \subs \{C \subs V:|C| \leq k\}$, and prove the following. 

\begin{lemma} \label{l:CK}
Let $\CC$ be a family of sets of size at most $k$ each on a groundset $V$ of size~$n$. 
Then there exists a $\CC$-resilient family $\Sa$ of  $O(k^2 \ln (n|\CC|))$ subsets of $V$ of size $r=\f{n-k}{k+1}$ each.
Furthermore, assigning to each set in $\Aa = \{S \subs V: |S|=r\}$ probability $\Delta=1/{n-k-2 \choose r-2}$
and applying randomized rounding $4 \ln (n|\CC|)$ times gives such ${\cal S}$ w.h.p.
\end{lemma}
\begin{proof}
If $n \leq 3k + 1$, then the subsets of $V$ of size $2$ is a family as required of size $\f{3k(3k+1)}{2}$, 
so assume that $n \geq 3k + 2$. 

Let $\Aa = \{S \subs V: |S|=r\}$ and $\Bb=\{(\{s,t\},C): s,t \in V, C \in \CC\}$.
Define a bipartite graph with sides $\Aa,\Bb$ by connecting $S \in \Aa$ 
to $(\{s,t\},C) \in \Bb$ if $s,t \in S$ and $C \cap S=\empt$;
in this case we will say that {\bf $S$ covers $(\{s,t\},C)$}.
This defines an instance of the {\sc Set Cover} problem,
where $\Aa$ are the sets and $\Bb$ are the elements. 
The lemma says that there exists a cover $\Sa \subs \Aa$ of $\Bb$ that has size $|\Sa|=O(k^2 \log (n|\CC|))$.

A {\bf fractional cover} of $\Bb$ is a function $h:\Aa \longrightarrow [0,1]$ such that 
$$\sum\{h(S):S \in \Aa \mbox{ covers } (\{s,t\},C)\} \geq 1 \ \ \ \ \  \forall (\{s,t\},C) \in \Bb \ .$$
The {\bf value} of a fractional cover $h$ is $\sum_{S \in \Aa}h(S)$.
It is known that if there is a fractional cover of value $\tau$, then there is a cover of size $\tau (1+\ln |\Bb|)$.
We have $|\Bb|=\f{n(n-1)}{2}|\CC|$, hence $\ln |\Bb| \leq 2\ln (n|\CC|) - \ln 2$ and $\lceil 2\ln |\Bb| \rceil \leq 4\ln (n|\CC|)$.

Our next goal is to show that there is a fractional cover of value $O(k^2)$.
We have $|\Aa|={n \choose r}$.
The number of sets in $\Aa$ that cover a given member $(\{s,t\},C) \in \Bb$
is $\Delta={n-k-2 \choose r-2}$, 
which is the number of choices of a set $S \sem \{s,t\}$ of size $r-2$ from 
the set $V \sem (C \cup \{s,t\})$ of size $n-k-2$.
Defining $h(S) = 1/\Delta$ for all $S \in \Aa$ gives a fractional cover of value $|\Aa|/\Delta$. Denote $m=n-k$. Then:   
$$
\f{|\Aa|}{\Delta} 
= \f{{n \choose r}}{{m-2 \choose r-2}} 
= \f{m(m-1)}{r(r-1)} \cdot \f{n!}{(n-r)!} \cdot \f{(m-r)!}{m!}   
\leq \f{m^2}{(r-1)^2} \prod_{i=1}^{r} \f{n-i+1}{m-i+1} \ .
$$
Note that for $1 \leq i \leq r$ we have 
$\f{n-i+1}{m-i+1} = 1+ \f{n-m}{m-i+1} \leq 1+\f{k}{n-k-r}$.
Let us choose $r$ such that $\f{k}{n-k-r}=\f{1}{r}$, so $r=\f{n-k}{k+1}$;
assume that $r$ is an integer, as adjustment to floors and ceilings 
only affects by a small amount the constant hidden in the $O(\cdot)$ term.
Since ${(1+1/r)}^r \leq e$ we obtain
$$
\prod_{i=1}^{r} \f{n-i+1}{m-i+1} \leq {\left(1+\f{1}{r}\right)}^r  \leq e \ .
$$
Since we assume that $n \geq 3k+2$, we have $\f{n-k}{k+1} \geq 2$ and thus $\f{m}{r-1} \leq 2(k+1)$.
Consequently, we get that 
$\f{|\Aa|}{\Delta} \cdot (1+\ln |\Bb|) =O(k^2 \ln (n|\CC|))$.
This implies that a standard greedy algorithm for {\sc Set Cover}, 
produces the required family of size $O\left(k^2 \ln n|\CC| \right)$. 
There is a difficulty to implement this algorithm in time polynomial in $n$ (unless $r=\f{n-k}{k+1}$ is a constant),
since $|\Aa|={n \choose r}$ may not be polynomial in $n$. 
Thus we use a randomized algorithm for {\sc Set Cover}, by rounding each entry to $1$ with probability
determined by our fractional cover.
It is known that repeating this rounding $2\lceil \ln |\Bb| \rceil \leq 4 \ln (n|\CC|)$ times gives a cover w.h.p.,
and clearly its expected size is $2\lceil \ln |\Bb| \rceil$ times the value of the fractional hitting set.
In our case, $h(S)=1/\Delta=1/{n-k-2 \choose r-2}$ for all $S \in \Aa$.
Thus we just need to assign to each set in $\Aa$ probability $1/\Delta$,
and apply randomized rounding $4 \ln (n|\CC|)$ times. 
\end{proof}

Applying Lemma~\ref{l:CK} on the family $\CC$ as in as in Lemma~\ref{l:list}, that has size $|\CC| \leq k(k+2)n$, 
we get that that there exists a $\CC$-resilient family $\Sa$ of  $O(k^2 \ln (n|\CC|))=O(k^2 \ln n)$ subsets of $V$ 
of size $r \approx \f{n-k}{k+1}$ each.
On the other hand, if $|\CC|$ is the family of all subsets of $V$ of size $k$, 
then $|\CC|={n \choose k} < (ne/k)^k$ and we get the bound $O(k^2 \ln (n|\CC|))=O(k^3 \ln n)$ of Chuzhoy and Khanna \cite{CK}. 


\end{document}